\documentclass[journal]{IEEEtran}

\usepackage{amsthm}
\usepackage{amssymb}
\usepackage{amsfonts}
\usepackage{amsfonts,subfigure,multicol,color,verbatim,
graphicx,cite,epsfig,amssymb,amsmath,cases,bm,algorithm,
algorithmic,xcolor,multirow,array}
\usepackage{multirow}
\usepackage{multicol}
\usepackage{supertabular}
\usepackage{array}
\usepackage{colortbl}
\usepackage{longtable}
\usepackage{subfigure}
\usepackage{dsfont}

\usepackage{epstopdf,booktabs}

\newtheorem{theorem}{Theorem}

\newtheorem{proposition}[theorem]{Proposition}

\begin{document}

\title{Deep Reinforcement Learning Based Mode Selection and Resource Management for Green Fog Radio Access Networks}

\author{Yaohua~Sun, \IEEEmembership{Student Member, IEEE}, Mugen~Peng, \IEEEmembership{Senior Member, IEEE}, \\ and Shiwen~Mao, \IEEEmembership{Senior Member, IEEE}
\thanks{Yaohua~Sun (e-mail: sunyaohua@bupt.edu.cn) and Mugen~Peng (e-mail: pmg@bupt.edu.cn) are with the Key Laboratory of Universal Wireless Communications (Ministry of Education), Beijing University of Posts and Telecommunications, Beijing, China. Shiwen Mao (smao@ieee.org) is with the Department of Electrical and Computer Engineering, Auburn University, Auburn, AL 36849-5201, USA. \emph{(Corresponding author: Mugen Peng)}}
}
%\author{Yaohua~Sun, Mugen~Peng, \IEEEmembership{Senior Member, IEEE}, and H. Vincent Poor, \IEEEmembership{Fellow, IEEE}
%\thanks{Yaohua~Sun (e-mail: sunyaohua@bupt.edu.cn) and Mugen~Peng (e-mail: pmg@bupt.edu.cn) are with the Key Laboratory of Universal Wireless Communications (Ministry of Education), Beijing University of Posts and Telecommunications, Beijing, China. H. Vincent Poor (poor@princeton.edu) is with the Department of Electrical Engineering, Princeton University, Princeton, NJ, USA.}}

\maketitle

\begin{abstract}
Fog radio access networks (F-RANs) are seen as potential architectures to support services of internet of things by leveraging edge caching and edge computing. However, current works studying resource management in F-RANs mainly consider a static system with only one communication mode. Given network dynamics, resource diversity, and the coupling of resource management with mode selection, resource management in F-RANs becomes very challenging. Motivated by the recent development of artificial intelligence, a deep reinforcement learning (DRL) based joint mode selection and resource management approach is proposed. Each user equipment (UE) can operate either in cloud RAN (C-RAN) mode or in device-to-device mode, and the resource managed includes both radio resource and computing resource. The core idea is that the network controller makes intelligent decisions on UE communication modes and processors' on-off states with precoding for UEs in C-RAN mode optimized subsequently, aiming at minimizing long-term system power consumption under the dynamics of edge cache states. By simulations, the impacts of several parameters, such as learning rate and edge caching service capability, on system performance are demonstrated, and meanwhile the proposal is compared with other different schemes to show its effectiveness. Moreover, transfer learning is integrated with DRL to accelerate learning process.

%At each decision step, the DRL based controller first intelligently determines the communication modes and processors' on-off states, and then
%precoding is optimized for UEs in cloud RAN mode

\end{abstract}

%\newpage


\section{Introduction}

%%引出c-ran，介绍c-ran的发展
%With the explosive increase in mobile data traffic, operators
%have to continuously improve network performance~\cite{survey}. As a promising solution,
%the cloud radio access network (C-RAN) has been proposed to enhance
%spectral efficiency (SE) and energy efficiency (EE)
%as well as lower operating expenditure and capital expenditure~\cite{net}.
%In C-RANs, the baseband processing and upper-layer functionalities
%are migrated to the baseband unit (BBU) pool, which connects with distributed remote radio heads (RRHs) via fronthaul.
%Many studies have been conducted on
%C-RANs, in terms of computing resource optimization~\cite{yan},
%performance analysis~\cite{si}, system cost minimization~\cite{shi}, and so on.
%
%%channel estimation~\cite{xie}

As a promising architecture, the fog radio access network (F-RAN) can well support the future services of
internet of things (IoT) with the help of edge caching and edge computing~\cite{fog,meng}.
These services include patient health monitoring~\cite{smart}, services with low latency~\cite{low},
large scale IoT data analytics~\cite{large}, and so on.
In F-RANs, each user equipment (UE) can potentially operate in different communication modes including
cloud RAN (C-RAN) mode, fog radio access point (FAP) mode, device-to-device (D2D) mode, and so on.
In C-RAN mode, UEs are served by multiple cooperative remote radio heads (RRHs), benefited from
centralized signal processing and resource allocation, while UEs in FAP mode and D2D mode
are served locally by FAPs and UEs equipped with cache, respectively.
Recently, many studies have been conducted on F-RANs, in terms of
performance analysis~\cite{tony}, radio resource allocation~\cite{tao}, the joint design of cloud and edge
processing~\cite{simo}, the impact of cache size~\cite{interchannel}, and so on.

Although significant progress has been achieved, resource management in F-RANs still
needs further investigation. Compared with resource management in traditional wireless networks,
communication mode selection should be addressed as well due to the coupling with resource management,
and meanwhile the dynamics of edge caching complicate the network environment,
which both lead to a more challenging problem.
Specifically, from the perspective of optimization, communication mode selection problem is
usually NP-hard~\cite{guanding}. To solve the problem, classical algorithms like
branch and bound and particle swarm can be adopted.
Nevertheless, considering the network dynamics, communication modes of UEs need to be
frequently updated, which makes algorithms with high complexity less applicable.

On the other hand, owing to the great development of fast and massively parallel graphical processing units
as well as the explosive growth of data, deep learning has attracted a lot of attention and is
widely adopted in speech recognition, image recognition, localization~\cite{xuyumag,xvyu}, and so on. To help the computer
learn environment from high-dimensional raw input data and make intelligent decisions,
the author in~\cite{minh} proposes to combine deep learning with reinforcement learning, and the proposal is known as
deep reinforcement learning (DRL). In DRL, the deep neural network (DNN) adopted as the Q-function approximator
is called deep Q network (DQN). Using replay memory and target DQN, DRL algorithm can realize stable training.

Totally speaking, applying DRL to wireless networks has the following considerable advantages.
First, a DNN with a moderate size can finish the prediction given the input in almost real time since
only a small number of simple operations is required for forward passing~\cite{sun}. This
facilitates a DRL agent to make a quick control decision on networks
based on the Q values output by the DQN.
Second, the powerful representation capabilities of DNNs allows the DRL agent to learn directly from
the raw collected network data with high dimension instead of manual inputs.
Third, by distributing computation across multiple
machines and multiple cores, the time to train DQN can be greatly
reduced~\cite{stan}. Fourth, the DRL agent aims at optimizing a long-term performance, considering the impact of actions
on future reward/cost. The fifth advantage of DRL is that it is a model-free approach, and hence
does not rely on a specific system model that can be based on some ideal assumptions. At last,
it is convenient for DRL based schemes to consider the cost incurred by system state transitions.

Motivated by the benefits of DRL, a DRL based joint mode selection and resource management approach is proposed,
aiming at minimizing the long-term F-RAN system power consumption. Using DRL, the controller can quickly
control the communication modes of UEs and the on-off states of processors in the cloud facing with
the dynamics of edge caching states. After the controller makes the decision,
precoding for UEs in C-RAN mode is optimized subsequently with quality of service (QoS) constraints
and the computing capability constraint in the cloud.

%
%%介绍本文是如何做的
%
\subsection{Related Work and Challenges}
%
%%介绍Green c-ran相关论文，10篇左右
%
Up to now, some attention has been paid to radio resource management in fog radio access networks.
In~\cite{tao}, a cloud radio access network with each RRH equipped with a cache is investigated,
and a content-centric beamforming design is presented, where a group of
users requesting the same content is served by a RRH cluster. A
mixed-integer nonlinear programming problem is formulated to minimize the weighted
sum of backhaul cost and transmit power under the QoS constraint of each user group.
While in~\cite{ada}, a similar network scenario is considered, aiming at minimizing
the total system power consumption by RRH selection and load balancing.
Specifically, the RRH operation power incurred by circuits and cooling system is included and backhaul capacity
constraint is involved.
The author in~\cite{sparse} goes one step further by jointly optimizing RRH selection, data assignment,
and multicast beamforming. The data assignment refers to whether the content
requested by a user group is delivered to a RRH via backhaul, which is not handled in~\cite{tao} and~\cite{ada}.
To solve the NP-hard network power consumption minimization problem, which consists of RRH power consumption and backhaul power consumption,
a generalized layered group sparse beamforming modeling framework is proposed.
Different from previous works that mainly optimize network cost or system power consumption, the author in~\cite{zheng}
tries to minimize the total content delivery latency of the network, which is the sum of wireless transmission latency and
backhaul latency caused by downloading the uncached contents. Due to the fractional form and $l$-0 norm in the objective
function, the formulated problem is non-convex, which is then decomposed into beamforming design problem and
data assignment problem.

Although the proposals in the above works achieve good performance, only C-RAN mode is taken into account.
When a UE in F-RANs is allowed to operate in different communication modes, the problem of user communication mode selection
should be handled, which is the key to gaining the benefits of F-RANs~\cite{survey}.
In~\cite{xiang}, the author investigates a joint mode selection and resource allocation problem in a downlink F-RAN,
and particle swarm optimization is utilized to optimize
user communication modes. Other approaches to mode selection problems include branch and bound~\cite{guanding} as well as Tabu search~\cite{japan}.
However, these optimization methods can induce high computational complexity.
While in~\cite{yan}, evolutionary game is adopted to model the interaction of users for mode selection,
in which the payoff of each user involves both the ergodic rate under a certain
mode and the delay cost. Then, an algorithm based on replicator dynamics is proposed to achieve
evolutionary equilibrium. Nevertheless, the proposed algorithm can get only the proportion of
users selecting each communication mode, and therefore accurate communication mode control
can not be realized.

Moreover, the works~\cite{tao,ada,sparse,zheng,xiang} research resource management problems under a
static environment where content availability at each cache is unchanged.
This assumption is reasonable for content delivery via FAPs or RRHs with a cache, since cached contents are
usually updated on a large time scale, and meanwhile
FAPs and RRHs have stable power supplies to keep the normal operations of their caches.
On the contrary, for content delivery via D2D transmission, cache state dynamics should be taken into account.
That is the local availability of the content requested by a UE at the cache of its paired UE can easily change with time,
incurred by the autonomous and frequent cache update behavior of the UE holders, the dynamic battery level of the paired UE,
user time-varying content requests, and so on.
These dynamics will make mode selection algorithms with high complexity inapplicable.
Even worse, the existence of interference between active D2D links and UEs in C-RAN mode complicates the wireless environment as well.

Fortunately, DRL, as an emerging approach to complicated control problems, has
the potential to provide efficient solutions for wireless network design.
In~\cite{yu}, a DRL based communication link scheduling algorithm is developed for a cache-enabled opportunistic
interference alignment wireless network. Markov process is used to model the network dynamics including the dynamics
of cache states at the transmitter side and channel state information (CSI).
To extract features from the high dimensional input composed of CSI and
cache states, a DNN with several convolutional layers is used to learn the state representation.
In~\cite{wen}, DRL is applied in mobility management, and a convolutional NN and a
recurrent NN are responsible for feature extraction from the Received
Signal Strength Indicator. The performance is evaluated on a practical testbed in
a wireless local area network, and significant throughput improvement is observed.
While in~\cite{tang}, the author revisits the power consumption minimization problem in
C-RANs using DRL to control the activation of RRHs, where the power consumption
caused by the RRH on-off state transition is considered as well.
In addition, the author in~\cite{wicache} shows that DRL based on Wolpertinger architecture is effective in cache management.
Specifically, the request frequencies of each file over different time durations and the current file requests
from users constitute the input state, and the action decides whether to cache the requested
content.

\subsection{Contributions and Organization}

In this paper, a network power consumption minimization problem for a downlink F-RAN is studied.
Different from~\cite{tao,ada,sparse,tang}, the power consumption induced by the running processors in the cloud
for centralized signal processing is included as well.
Owing to the caching capability of D2D transmitters, UEs can acquire the desired contents locally without
accessing RRHs. This can help traffic offloading which alleviates the burden of fronthaul on one hand and on the other hand
allows turning off some processors in the cloud to save energy since less computing resource is needed to support
less number of UEs.
Facing with the dynamic cache states at D2D transmitters and the interference
between UEs in the same communication mode, a DRL based approach is proposed to help the network controller
learn the environment from raw collected data and make intelligent and fast decisions on network operations to reduce system power consumption.
\emph{As far as we know, our paper is the first work to adopt DRL to solve joint communication mode selection and resource management problem
taking the dynamics of edge cache states into account to achieve a green F-RAN.}
The main contributions of the paper are:
\begin{itemize}
\item An energy minimization problem in a downlink F-RAN with two potential communication modes, i.e., C-RAN mode and D2D mode, is investigated. To make the system model more practical, the dynamics of cache states at D2D transmitters are considered, which are modeled by Markov process, and the power consumption caused by processors in the cloud is taken into account. Based on the system model, a Markov decision problem is formulated, where the network controller aims at minimizing the long-term system power consumption by controlling UE communication modes and processors' on-off states at each decision step with the precoding for UEs in C-RAN mode optimized subsequently.
\item For the precoding optimization under given UE communication modes and processors' on-off states, the corresponding problem is formulated as a RRH transmission power
minimization problem under per-UE QoS constraints, per-RRH transmission power constraints, and the computing resource constraint in the cloud, which is solved by an iterative algorithm based on $l$-0 norm approximation. Then, a DRL based approach is proposed with UE communication modes, edge cache states, and processors' on-off states as input to select actions for communication mode and processor state control. After precoding optimization, the negative of system power consumption is determined and fed back to the controller as the reward, based on which the controller updates DQN.
\item The impacts of important learning parameters and edge caching service capability on system performance are illustrated,
and the proposal is compared with other several communication mode and processor state control schemes including Q-learning and random control.
Furthermore, the effectiveness of integrating transfer learning with DRL to accelerate the training process in a new but similar environment is demonstrated.
\end{itemize}

The remainder of this paper is organized as follows. Section
II describes the downlink F-RAN model.
Section III formulates the concerned energy minimization problem,
and the DRL based approach is specified in Section IV.
Simulation results are illustrated in Section
V, followed by the conclusion in Section VI. %For convenience, some important notations are listed in Table I.

\section{System Model}

%-----------把processor通过光纤互联写上------------
The discussed downlink F-RAN system is shown in Fig. 1, which consists of one cloud, multiple RRHs,
multiple UEs with their paired D2D transmitters. The cloud contains multiple processors of heterogenous computing capabilities which
are connected with each other via fiber links to achieve computing resource sharing, and the computing capability of processor $n$ is characterized
by $D_n$ whose unit is million operations per time slot (MOPTS)~\cite{liao}.
For each processor in the cloud, it has two states, i.e., on state and off state, which are indicated by $s_n^{processor} = 1$
and $s_n^{processor} = 0$, respectively.
Meanwhile, content servers provide large-scale caching capability, and the controller is used for network control like resource management.
The set of processors, RRHs, and UEs are denoted by
${\cal N} = {1,2,...,N}$, ${\cal K} = {1,2,...,K}$, and ${\cal M} = {1,2,...,M}$, respectively.
Each RRH is equipped with $L$ antennas, and each UE is with one antenna.
In addition, RRHs communicate with the cloud via high-bandwidth fronthaul.

The paired D2D transmitter for each UE is chosen by comprehensively considering the social tie and the physical condition as per in~\cite{social}.
In the considered scenario, each UE can operate either in D2D mode or C-RAN mode which are denoted by a 0-1 indicator $s_m^{mode}$. Specifically,
$s_m^{mode}=1$ means UE $m$ operates in D2D mode, while
$s_m^{mode}=0$ means that UE $m$ is served by RRHs.
Moreover, suppose that the D2D transmission does not interfere the RRH transmission by operating in different frequency bands,
and all the UEs in the same communication mode share the same frequency band, hence interfering with each other.
Finally, the high power node (HPN) with wide coverage is responsible for delivering control signalling and exchanges control information with the controller via backhaul~\cite{hcran}.
In the following, models for communication, computing, caching, and system energy consumption are elaborated.

\begin{figure}[!t]
\center
\includegraphics[width=0.47\textwidth]{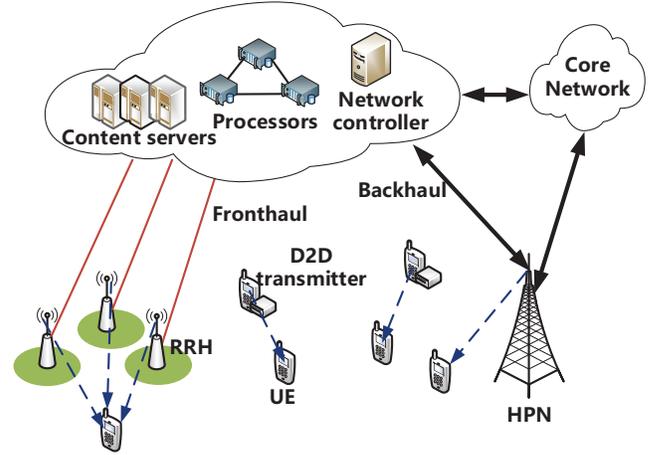}
\caption{A downlink F-RAN system.} \label{cran}
\end{figure}

%-----------介绍信道建模----------在下面公式中加入时间t
\subsection{Communication Model}

By the collaborative transmission of RRHs, the received symbol of UE $m$ in C-RAN mode is given by
\begin{align}
{y_m^C} =&\; \sum\limits_{k \in \cal K} {{\bf{h}}_{m,k}^H{{\bf{v}}_{m,k}}{x_m}}  + \nonumber \\
&\; \sum\limits_{m' \in {\cal M},m' \ne m,{s_{m'}^{mode}} = 0} {\sum\limits_{k \in \cal K} {{\bf{h}}_{m,k}^H{{\bf{v}}_{m',k}}{x_{m'}}} }  + {z_m},
\end{align}
where ${x_m}$ is the message of UE $m$, ${\bf{h}}_{m,k}^{}$ is the channel vector between RRH $k$ and UE $m$,
${{\bf{v}}_{{m,k}}}$ is the precoding vector of RRH $k$ for UE $m$, and $z_m$ is the noise which follows the distribution of ${\cal CN}\left( {0,{\sigma ^2}} \right)$.
Then the data rate achieved by UE $m$ is given by
\begin{equation}
{R_m^C} = \log \left( {1 + \frac{{{{\left| {\sum\limits_{k \in {\cal K}} {{\bf{h}}_{m,k}^H{{\bf{v}}_{m,k}}} } \right|}^2}}}{{{\sum\limits_{m' \in {\cal M},m' \ne m,{s_{m'}^{mode}} = 0} {{{\left| {\sum\limits_{k \in {\cal K}} {{\bf{h}}_{m,k}^H{{\bf{v}}_{m',k}}} } \right|}^2}} } + {\sigma ^2}}}} \right).
\end{equation}

For UE $m$ in D2D mode, it is assumed that the D2D transmitter transmits at a constant power level, and the received symbol of UE $m$ is given by
\begin{equation}
{y_m^D} = \sqrt {p_m} h_{m}{x_m} + \sum\limits_{m' \in {\cal M},m' \ne m,{s_{m'}^{mode}} = 1} {\sqrt {{p_{m'}}} h_{m,m'}{x_{m'}} + {z_{m}}},
\end{equation}
where ${p_{m}}$ is the transmit power of the D2D transmitter paired with UE $m$,
$h_{m}$ is the channel coefficient between UE $m$ and its D2D transmitter,
$h_{m,m'}$ is the channel coefficient between the D2D transmitter of UE $m'$ and UE $m$.
Then the data rate achieved by UE $m$ in D2D mode is given by
\begin{equation}
R_{\rm{m}}^{(D)} = \log (1 + \frac{{{p_m}|{h_m}{|^2}}}{{\sum\nolimits_{m' \in {\cal M},m' \ne m,{s_{m'}^{mode}} = 1} {{p_{m'}}|{h_{m,m'}}{|^2} + {\sigma ^2}} }}).
\end{equation}

\subsection{Computing Model}
To finish the basedband processing and generate the transmitted signals for RRHs, computing resource provision in the cloud plays a key role, whose
model follows that in~\cite{liao}. Specifically, the computing resource consumed by coding and modulation for UE $m$
is given by
\begin{equation}
{D_{m,1}} = \beta {R_m},
\end{equation}
where ${R_m}$ is the data rate of UE $m$.
Meanwhile, the computing resource consumed by calculating the transmit signal for UE $m$ depends on the number of non-zeros elements in its network wide precoding vector
$\bf{v}_m$, which is modeled as
\begin{equation}
{D_{m,2}} = \alpha {\left\| {{{\bf v}_m}} \right\|_0}.
\end{equation}
Then, the computing resource consumption for the whole system is calculated as
\begin{align}
{D_{system}} =&\; \sum\limits_{m,{{s_{m}^{mode}}} = 0} {\left( {{D_{m,1}} + {D_{m,2}}} \right)} \nonumber \\
 =&\; \beta\sum\limits_{m,{{s_{m}^{mode}}} = 0} { {R_m}}  + \alpha\sum\limits_{m,{{s_{m}^{mode}}} = 0} { {{\left\| {{{\bf{v}}_m}} \right\|}_0}}.
\end{align}
Note that only the UEs accessing RRHs consume computing resource, and hence computing resource needed may decrease
by serving UEs locally via D2D communication, which further allows turning off some processors to save energy.
Moreover, it should be highlighted that additional constant terms can be added to (7) to account for the computing resource consumed by other baseband operations,
which has no impact on our proposal.

\subsection{Caching Model}

We define the value of the cache state at a D2D transmitter is \textbf{True} only when
the requested content is cached in the D2D transmitter and the transmitter's battery level is high enough so that
the holder is willing to share contents, and the cache state is \textbf{False} otherwise.
Note that the cache state at each D2D transmitter can be highly dynamic due to
the following reasons.
First, although UEs are paired based on their social ties, this can not imply
the content requested by a UE must be cached by its partner whose cached contents can be frequently updated by the device holder based on its own interest.
Second, the UE battery level dynamically changes with time, and the user content request is time-varying.

To characterize the dynamics of cache states, Markov process is adopted as per~\cite{yu} with the
probability transition matrix given by
\begin{equation}
P{r_{cache}} = \left[ {\begin{array}{*{20}{c}}
{P{r_{True,True}}}&{P{r_{True,False}}}\\
{P{r_{False,True}}}&{P{r_{False,False}}}
\end{array}} \right],
\end{equation}
where ${{{Pr }_{True,False}}}$ denotes the transition probability of the cache state at a D2D transmitter from \textbf{True} to \textbf{False}.

\subsection{Energy Consumption Model}

Totally speaking, the energy consumption in the concerned downlink F-RAN includes the energy consumption incurred by the running processors, fronthaul transmission,
and wireless transmission. First, according to~\cite{vm}, the energy consumed by processor $n$ in Watts is given by
\begin{equation}
E_n^{processor} = s_n^{processor}\mu D_n^3,
\end{equation}
where $\mu$ is a parameter depending on the structure of the processor.
Second, the wireless transmission power for UE $m$ is as follows.
\begin{equation}
E_m^{wireless} = \left( {1 - {{s_{m}^{mode}}}} \right)\frac{1}{{{\eta _1}}}\left\| {{{\bf{v}}_m}} \right\|_2^2 + {{s_{m}^{mode}}}\frac{1}{{{\eta _2}}}{P_m},
\end{equation}
which is the transmission power when UE $m$ is either served by RRHs or served by its paired D2D transmitter.
${\eta _{\rm{1}}}$ and ${\eta _{\rm{2}}}$ are the efficiency of the power amplifier at each RRH and at each UE, respectively~\cite{vm}.
Third, for the fronthaul energy consumption corresponding to UE $m$, it is simply modeled as
\begin{equation}
E_m^{fronthaul} = \left( {1 - {{s_{m}^{mode}}}} \right)P_m^{front},
\end{equation}
with $P_m^{front}$ a constant representing the energy consumption for delivering the processed signal
of UE $m$ to its associated RRHs via fronthaul~\cite{ada}.
Then, the energy consumption of the whole system is given by
\begin{equation}\label{system}%把权重补上
{E_{system}} = \sum\limits_n {E_n^{processor}}  + \sum\limits_m {E_m^{fronthaul}}  + \sum\limits_m {E_m^{wireless}}.
\end{equation}

It should be noted that the modeling of the caching state using a Markov process motivates the adoption of the Markov decision process (MDP)
to formulate our concerned problem. In addition, since our aim is to achieve a green F-RAN under user QoS and computing resource constraints
, the reward setting of the MDP will be closely related to the data rate, computing, and energy consumption models.

\section{Problem Formulation and Decoupling}

%-------------mdp----------reinforcement learning----------deep reinforcement learning（重点突出deep reinforcement learning相对于传统q learning的优势，非线性近似时会有问题之外，还有什么缺点？？）-------------

In this section, an optimization problem aiming at minimizing the energy consumption in a downlink F-RAN is formulated from an MDP perspective.
Specifically, the problem is decoupled into a joint control problem of processors' on-off states and UEs' communication modes and a precoding design problem under the computing resource constraint.

\subsection{The Basics of MDP}

%----------把BBU pool改成cloud
From the energy consumption model, it can be seen that the cache state for each UE plays a key role in the following ways.
On one hand, the cache state will influence the set of UEs that can potentially exploit D2D communication, which
directly influences the energy consumption incurred by fronthaul and wireless transmission. On the other hand,
since UEs served by local caching do not consume computing resource in the cloud anymore,
there is a chance to turn off some processors to save energy.
Facing with the dynamics of cache state for each UE pair, it is natural to formulate the energy minimization problem
from an MDP perspective.

MDP provides a formalism for reasoning about planning and acting in the face of uncertainty, which can be
defined using a tuple $\left( {{\cal S},{\cal A},\left\{ {{{\Pr }_{sa}}\left(  \cdot  \right)} \right\},U} \right)$.
$\cal S$ is the set of possible states, $\cal A$ is the set of available actions, $\left\{ {{{\Pr }_{sa}}\left(  \cdot  \right)} \right\}$ gives
the transition probabilities to each state if action $a$ is taken in state $s$, and $U$ is the reward function. The process of MPD is described as follows. At an initial state $s_0$, the agent takes an action $a_0$. Then
the state of the system transits to the next state $s_1$ according to the transition probabilities $\left\{ {{{\Pr }_{{s_0}{a_0}}}\left( \cdot  \right)} \right\}$,
and the agent receives a reward $U_0$. With the process continuing, a state sequence
${s_0},{s_1},{s_2}, \ldots$ is generated. The agent in an MPD aims to maximize a discounted accumulative reward when starting in state $s$,
which is called state-value function and defined as
\begin{equation}
{V^\pi }\left( s \right) = \mathds{E}\left[ {\sum\limits_{t = 0}^\infty  {{\gamma ^t}U_t} \left| {{s_0} = s,\pi } \right.} \right],
\end{equation}
where ${{U_t}}$ is the reward received at decision step $t$, $\gamma $ is a discount factor adjusting the
effect of future rewards to the current decisions, and
the policy $\pi $ is a mapping from state $s$ to a probability distribution over actions that the agent can take in state $s$.

The optimal state-value function is given by
\begin{equation}
{V^*}\left( s \right) = \mathop {\max }\limits_\pi  {V^\pi }\left( s \right).
\end{equation}
Then, if ${V^*}\left( s \right)$ is available, the optimal policy ${\pi ^*}$ is determined as
\begin{equation}
{\pi ^*}\left( {{s_t}} \right) = \arg \mathop {\max }\limits_{a \in \cal A} \bar U_{{s_t}}^a + \sum\limits_{{s_{t + 1}}} {{{\Pr }_{{s_t}a}}\left( {{s_{t + 1}}} \right){V^*}\left( {{s_{t + 1}}} \right)},
\end{equation}
where $\bar U_{{s_t}}^a$ is the expected reward by taking action $a$ at state $s_t$.
To calculate ${V^*}\left( s \right)$, the value-iteration algorithm can be adopted. However,
since the transition probabilities ${\left\{ {{{\Pr }_{sa}}\left(  \cdot  \right)} \right\}}$ are not easy to acquire
in many practical problems, reinforcement learning algorithms, especially Q-learning, are widely adopted to
handle MDP problems, for which the state space, explicit transition probabilities, and the reward function are not essential~\cite{sutton}.

In Q-learning, the Q function is defined as
\begin{equation}
{Q^\pi }\left( {s,a} \right) = \mathds{E}\left[ {\sum\limits_{t = 0}^\infty  {{\gamma ^t}U_t} \left| {{s_0} = s,{a_{\rm{0}}} = a,\pi } \right.} \right],
\end{equation}
which is the expected accumulative reward when starting from state $s$ with action $a$ and then following policy $\pi$.
Similarly, we can define the optimal Q function as
\begin{equation}
{Q^*}\left( {s,a} \right) = \mathop {\max }\limits_\pi  {Q^\pi }\left( {s,a} \right).
\end{equation}
Q-learning is ensured to reach the optimal Q values under certain conditions~\cite{sutton}, which is executed iteratively according to
\begin{equation}\label{Q}
{Q_{t + 1}}\left( {s,a} \right) = \left( {1 - \alpha } \right){Q_t}\left( {s,a} \right) + \alpha \left[ {{U_t} + \gamma \mathop {\max }\limits_{a'} {Q_t}\left( {{s_{t+1}},a'} \right)} \right],
\end{equation}
where $\alpha $ is the learning rate.
Once ${Q^{\rm{*}}}\left( {s,a} \right)$ for each state-action pair is achieved, the optimal policy can be determined as
\begin{equation}
{\pi ^*}\left( s \right) = \arg \mathop {\max }\limits_{a \in \cal A} {Q^{\rm{*}}}\left( {s,a} \right).
\end{equation}
\subsection{Problem Formulation}

In this paper, the energy minimization problem for the considered downlink F-RAN is formulated as an MDP problem, where
the controller in the cloud tries to minimize long-term system energy consumption by controlling the
on-off states of processors, the communication mode of each UE, and optimizing precoding vectors for RRH transmission.
More formally, our concerned MPD problem is defined as follows.
\begin{itemize}
  \item \textbf{State space}: The state space $\cal S$ is defined as a set of tuples ${\cal S}{\rm{ = }}\left\{ {\left\{ {{{\bf s}^{processor}},{{\bf s}^{mode}},{{\bf s}^{cache}}} \right\}} \right\}$. ${{{\bf s}_{processor}}}$ is a vector representing the current on-off states of all the processors, where the $n$-th element
      is $s_n^{processor}$. ${{{\bf s}_{mode}}}$ is a vector representing the current communication modes of all the UEs, where the $m$-th element is $s_m^{mode}$.
      While ${{{\bf s}^{cache}}}$ is a vector consisting of the cache state at each D2D transmitter.
  \item \textbf{Action space}: The action space $\cal A$ is defined as a set of tuples ${{{\cal A} = }}\left\{ {\left\{ {{a_{processor}},{a_{mode}}} \right\}} \right\}$. ${{a_{processor}}}$ represents to turn on or turn off a certain processor, while ${{a_{mode}}}$ represents to change the communication mode of a certain UE. Note that the network controller controls the on-off state of only one processor and the communication mode of only one UE each time to reduce the number of actions~\cite{tang}.
      Moreover, the precoding design for RRH transmission is handled separately for the same reason~\cite{tang}.
  \item \textbf{Reward}: The immediate reward $U$ is taken as the negative of system energy consumption which is the sum of energy consumption incurred by the running processors, fronthaul transmission, and wireless transmission as defined in (\ref{system}). Hence, after communication mode and processor state control followed by the precoding optimization for UEs in C-RAN mode, the reward can be totally determined.
\end{itemize}

Note that due to the cache state dynamics,
the state after control can transit to an infeasible state.
In summary, three situations should be properly handled.
The first one is that the controller selects D2D mode for a UE, but the cache state at its paired UE after transition is \textbf{False}.
The second one is that the QoS of a UE in the C-RAN mode is not met due to too many sleeping processors, and the third one is that the QoS of a UE
in D2D mode is unsatisfied because of the strong interference among active D2D links.
To deal with these situations and always guarantee the QoS of UEs,
protecting operations will be performed.
Specifically, the UE with QoS violation will inform the HPN over the control channel, and then the HPN sends protecting operation information
to the controller that reactivates all the processors and switches each UE in D2D mode with QoS violation to RRHs.
In addition, the precoding for UEs in C-RAN mode will be re-optimized.

%The size of the state space for the controller is given by ${2^{N + 2*M}}$, which exponentially increases with the number of processors and UEs under control.
%Therefore, if Q-learning in (\ref{Q}) is applied, a huge table must be maintained to record the state-action values,
%and meanwhile it can cost long time to achieve a good control policy.
%Facing with these issues, function approximation can be adopted as an efficient approach aiming
%to generalize from examples of a function to construct an approximate of the entire function.
%Possible function approximator includes linear function, decision trees, and tile coding, which can only lead to
%a shallow approximation~\cite{drlsurvey}. In this paper, considering the powerful function approximation capability of DNNs,
%they are adopted to approximate the optimal Q-values, and the learned knowledge about the environment is stored in the form of NN parameters.
Motivated by the recent advances in artificial intelligence, DRL is utilized to control the on-off states of processors and the communication modes
of UEs. The details about the DRL based approach will be introduced in the next section.
After the controller takes an action using DRL, precoding is then optimized for RRH transmission,
which is formulated as the following optimization problem.
\begin{equation}
\begin{array}{*{20}{l}}
{\mathop {\min }\limits_{\left\{ {{{\bf{v}}_m}} \right\}} \sum\limits_{m,{s_m^{mode}} = 0} {\left\| {{{\bf{v}}_m}} \right\|_2^2} }\\
{\left( {a1} \right){R_m} \ge {R_{m,\min }},\forall m,}\\
{\left( {a2} \right)\sum\limits_{m,{s_m^{mode}} = 0} {\left\| {{{\bf{v}}_{m,k}}} \right\|_2^2}  \le {p_{\max }},\forall k,}\\
{\left( {a3} \right)\beta \sum\limits_{m,{s_m^{mode}} = 0} {{R_m}}  + \alpha\sum\limits_{m,{s_m^{mode}=0}} { {{\left\| {{{\bf{v}}_m}} \right\|}_0}} \le \sum\limits_n {s_n^{processor}{D_n}} },\\
\end{array}
\end{equation}
where ${{\bf{v}}_m}$ is the network wide precoding vector for UE $m$, the first constraint is to meet the QoS demand for each UE, the second constraint is the transmission power constraint for each RRH, while the last constraint
is the computing resource constraint in the cloud. Note that once the controller takes an action, the parameters
${\left\{ {{s_m^{mode}}} \right\}}$ and ${\left\{ {s_n^{processor}} \right\}}$ will be determined.

\section{DRL Based Mode Selection and Resource Management}

In this section, the precoding optimization given processors' on-off states and communication modes of UEs is handled first,
and then a DRL based algorithm is proposed to control the network facing with the dynamics of caching states at D2D
transmitters and complex radio environment.

\subsection{Precoding Design with the Computing Resource Constraint}

For problem (20), the main difficultly to solve it lies in the non-convex constraint ($a1$) as well as the $l$-0 norm and the sum rate in constraint ($a3$).
Fortunately, the QoS constraint ($a1$) can be transformed into a second order cone constraint by the phase rotation of precoding~\cite{vm}.
Moreover, the $l$-0 norm term in constraint ($a3$) can be approximated by re-weighted $l$-1 norm as per~\cite{liao} and~\cite{yuwei}.
Then, inspired by the proposal in~\cite{yuwei}, problem (20) can be solved iteratively and the problem for each iteration is as follows:
\begin{equation}
\begin{array}{*{20}{l}}
{\mathop {\min }\limits_{\left\{ {{{\bf{v}}_m}} \right\}} \sum\limits_{m,{s_m^{mode}} = 0} {\left\| {{{\bf{v}}_m}} \right\|_2^2} }\\
{\left( {e1} \right)\sqrt {\sum\limits_{m',{s_{m'}^{mode}=0}} {{{\left| {{\bf{h}}_{m}^H{{\bf{v}}_{m'}}} \right|}^2} + {\sigma ^2}} }  \le \sqrt {1 + \frac{1}{{{\gamma _m}}}} Re\left\{ {{\bf{h}}_m^H{{\bf{v}}_m}} \right\},\forall m,}\\
{\left( {e2} \right){\rm{Im}}\left\{ {{\bf{h}}_m^H{{\bf{v}}_m}} \right\} = 0,\forall m,}\\
{\left( {e3} \right)\sum\limits_{m,s_m^{mode} = 0} {\left\| {{{\bf{v}}_{m,k}}} \right\|_2^2}  \le {p_{\max }},\forall k,}\\
\left( {e4} \right)\beta\sum\limits_{m,s_m^{mode} = 0} { {{\tilde R}_m}}  + \nonumber \\
\;\;\;\;\;\;\; \alpha \sum\limits_{m,{s_m^{mode}}=0} {\sum\limits_k {\sum\limits_l {{\theta _{m,k,l}}\left| {{v_{m,k,l}}} \right|} } }  \le \sum\limits_n {s_n^{processor}{D_n}}, \\
\left( {e5} \right){v_{m,k,l}} = 0,if\;it\;is\;set\;to\;0\;in\;the\;last\;iteration,
\end{array}\tag{21}
\end{equation}
where ${\bf{h}}_m$ is the channel vector from all the RRHs to UE $m$, ${v_{m,k,l}}$ is the precoding of the $l$-th antenna of RRH $k$ for UE $m$, ${{\tilde R}_m}$ is the data rate of UE $m$ calculated by the precoding output by the last iteration, $\sum\limits_{m,{s_m^{mode}}=0} {\sum\limits_k {\sum\limits_l {{\theta _{m,k,l}}\left| {{v_{m,k,l}}} \right|} } }$ is the norm approximation of the term $\sum\limits_{m,s_m^{mode}=0} { {{\left\| {{{\bf{v}}_m}} \right\|}_0}}$ in constraint ($a3$) of problem (20).
${\theta _{m,k,l}}$ is updated as
\begin{equation}\label{wei}
{\theta _{m,k,l}} = \frac{1}{{\left| {{{\tilde v}_{m,k,l}}} \right| + \xi }},\;\forall m,\forall k,\forall l,\tag{22}
\end{equation}
with ${\tilde v_{m,k,l}}$ the precoding calculated by the last iteration and $\xi$ a small enough parameter.
Note that problem (21) is a convex optimization problem that can be efficiently solved by CVX~\cite{cvx},
and the proof of the convexity is given by the following proposition.
\begin{proposition}
Problem (21) is a convex optimization problem.
\begin{proof}
First, it has been shown that constraints ($e3$) and ($e5$) are convex in \cite{beam1}, and meanwhile,
the objective function as well as constraints ($e1$) and ($e2$) are also convex according to \cite{beam2}.
For the constraint ($e4$), it can be reformulated as the following inequality:
\begin{equation}
\begin{array}{l}
\sum\limits_{m,s_m^{mode} = 0} {\sum\limits_k {\sum\limits_l {{\theta _{m,k,l}}\left| {{v_{m,k,l}}} \right|} } }  \le \\
\frac{{\rm{1}}}{\alpha }\left( {\sum\limits_n {s_n^{processor}{D_n}} {\rm{ - }}\beta \sum\limits_{m,s_m^{mode} = 0} {{{\tilde R}_m}} } \right)
\end{array}\tag{23}
\end{equation},
where the right side is a constant and the left side is a convex reweighted $l$-1 norm \cite{beam3}.
Hence, it can be concluded that problem (21) is convex.
\end{proof}
\end{proposition}

The algorithm diagram is listed in Algorithm 1. First, the precoding is initialized, which can be got by solving a relaxed version of problem (20) without
considering the computing resource constraint.
Then, ${{\tilde R}_m}$ and the weight ${\theta _{m,k,l}}$ can be calculated, based on which problem (21) can be solved.
At the end of each iteration, ${v_{m,k,l}}$ with small enough $\left| {{{ v}_{m,k,l}}} \right|$ is set to 0, and one possible criteria is
comparing the value of $\left| {{{ v}_{m,k,l}}} \right|$ with $\xi$~\cite{ada}.
\begin{algorithm}[htbp]
\begin{algorithmic}[1]\caption{Precoding optimization under the computing resource constraint}
\STATE \textbf{Stage 1:}\\
The controller initializes precoding ${ v_{m,k,l}}, \forall m,\forall k,\forall l$, and then computes the weight $\theta _{m,k,l}$ and data rate ${{\tilde R}_m}$.
\STATE \textbf{Stage 2:}\\
Compute the optimal precoding by solving problem (21) using CVX;\\
Update weight $\theta _{m,k,l}$ based on the precoding result;\\
Let ${v_{m,k,l}}$ equal to 0 if $\left| {{{ v}_{m,k,l}}} \right|$
is less than a given small threshold.
\STATE \textbf{Stage 3:}\\
Repeat \textbf{Stage 2} until the RRH transmit power consumption converges.\\
\end{algorithmic}
\end{algorithm}
By updating $\theta _{m,k,l}$ iteratively, the above algorithm gradually sets ${v_{m,k,l}}=0$ for the UE-antenna link that has low transmit power~\cite{yuwei}. Meanwhile, utilizing the interior point method, the complexity of the above algorithm for each iteration is
${\cal O}\left({\left( {KLM} \right)^{3.5}}\right)$.

\subsection{DRL based Mode Selection and Resource Management}

After the precoding design under fixed UE communication modes and processors' on-off states is handled, the remaining task is to
find a way of reaching a good policy for the MDP formulated in Subsection B of Section III.
As introduced before, Q-learning is a widely adopted algorithm in the research of wireless networks for
network control without knowing the transition probability and reward function in advance.
However, Q-learning has three factors which can limit its application in future wireless networks.
First, traditional Q-learning stores the Q-values in a tabular form.
Second, to achieve the optimal policy,
Q-learning needs to revisit each state-action pair infinitely often~\cite{sutton}.
Third, the state for Q-learning
is often manually defined like in~\cite{bennis}.
These three characteristics will make Q-learning impractical when considering
large system state and action spaces~\cite{drlsurvey}.
While DRL proposed in~\cite{minh} can overcome these problems, and has the potential to
achieve better performance owing to the following facts.
First, DRL uses DQN to store learned Q-values in the form of connection weights between different layers.
Second, with the help of replay memory and generalization capability brought by
NNs, DRL can achieve good performance with less interactions with complex environments.
Third, with DQN, DRL can directly learn the representation from high dimensional raw network data,
and hence manual input is avoided.

Considering these benefits, the controller uses DRL to learn the control policy of UE communication modes and processors' on-off states
by interacting with the dynamic environment to minimize the discounted and accumulative system power consumption for $T$ decision steps,
and that is to maximize the long term reward given by $\mathds{E}\left[ {\sum\limits_{t = 0}^{T - 1} {{\gamma ^t}{U_t}} } \right]$.
The training procedure of the DRL is shown in Algorithm 2.
Specifically, given the current system state composed of UE communication modes, cache states at D2D transmitters, and processors' states,
the controller takes this state as input of DQN to output the Q-values $Q\left( {{s},{a},{\bf w}} \right)$ corresponding with each action.
Then, an action is selected based on $\varepsilon$ greedy scheme, and the operational states of a certain processor and a certain UE are
changed if needed. Afterward, the controller optimizes the precoding using Algorithm 1,
and the cache state at each D2D transmitter transits according to the transition matrix.
Once any QoS violation information from UEs is received by the HPN, the HPN will
help those UEs with unsatisfied QoS in D2D mode access the C-RAN and the controller will activate all the processors.
Next, this interaction is stored in the replay memory of the controller containing the state transition, the action,
and the negative of system power consumption which is the reward.
After several interactions, the controller will update DQN by training over a batch of interaction data
randomly sampled from the replay memory, intending to minimize the mean-squared-error between the target Q values and the predicted Q values of DQN.
In addition, every larger period, the controller will set the
weights of DQN to the target DQN.

\begin{algorithm}[htbp]
\begin{algorithmic}[1]\caption{DRL based communication mode selection and resource management}
\STATE \textbf{Stage 1:}\\
Randomly initialize a deep neural network as DQN with parameters ${\bf w}$ and make a copy of it to construct the target DQN with parameters $\hat {\bf w}={\bf{w}}$.
Then, the controller randomly selects actions for a period of time to store enough interaction
samples into the replay memory, each of which consists of the state transition, the action, and reward .\\
\STATE \textbf{Stage 2:}\\
\textbf{For} epoch $e=0,1,...,E-1$:\\
Initialize the beginning state $s_0$.\\
~\textbf{For} decision step $t=0,1,...,T-1$:\\
~~Generate a random number $x$ between 0 and 1.\\
~~\textbf{If} ${\rm{x}} \le \varepsilon$:\\
~~~The controller randomly selects an action.\\
~~\textbf{Else}:\\
~~~The controller selects the action $a_t$ as ${a_t} = \arg \;\mathop {\max }\limits_{a \in \cal A} Q\left( {{s_t},a,{\bf{w}}} \right)$.\\
~~\textbf{If end}.\\
~~The system state $s_t$ transits to $s_{t+1}$ according to the executed action $a_t$ and the cache state transition matrix.\\
~~Optimize the precoding for UEs in C-RAN mode by Algorithm 1.\\
~~\textbf{If} any UE reports a QoS violation to the HPN via the control channel, the HPN delivers the message to the controller via backhaul and
the controller then executes the protecting operation as specified in Subsection B of Section III.\\
~~\textbf{If end}.\\
~~The controller stores the reward $U_t$ which is the negative of the system power consumption together with $s_t$, $s_{t+1}$, and $a_t$ as an interaction sample
into the replay memory. When the capacity of the replay memory is full, the earliest sample is abandoned.\\
~~\textbf{If} the remainder when $t+1$ is divided by $T'$ is 0:\\
~~~Randomly fetch a mini-batch of interaction samples from the replay memory, and perform a single gradient update on
${\left( {{r_t} + \hat Q\left( {{s_{t + 1}},{a_t},\hat {\bf w}} \right) - Q\left( {{s_t},{a_t},{\bf w}} \right)} \right)^2}$
with respect to ${\bf{w}}$.\\
~~\textbf{If end}.\\
~~Periodically set the value of parameters ${\bf{w}}$ to $\hat {\bf w}$.\\
~\textbf{For end}.\\
\textbf{For end}.
\end{algorithmic}
\end{algorithm}

In addition to the proposal in~\cite{minh}, researchers have made some enhancements on DRL subsequently.
To more effectively reuse the experienced transitions in the replay memory,
prioritized replay is proposed. Moreover, double DRL is introduced to
overcome the optimistic Q-value estimation involved in the calculation of the target value,
while dueling DRL is proposed to effectively learn in the situation where the state value should be more cared about.
Furthermore, DRL with deep deterministic policy gradient is introduced to address the continuous control problem.
All these new DRL approaches take the advantage of the ideas of replay memory and target DQN in~\cite{minh},
and their specifications can be referred to~\cite{drlsurvey}.
Although only the proposal in~\cite{minh} is adopted for our communication mode selection and resource management in this paper,
these advances can be utilized as well, which does not affect the core idea and
the main conclusions of the paper.

\section{Simulation Results and Analysis}

\begin{figure}[!t]
\center
\includegraphics[width=3.3in, height=2.5in]{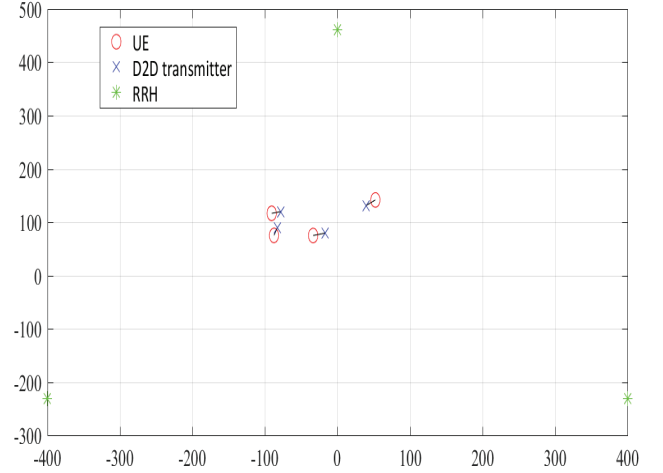}
\caption{The simulation scenario.}\label{location}
\end{figure}

The simulation scenario is illustrated in Fig. \ref{location} where the distance between each
pair of RRHs is 800 m, and four UEs are randomly distributed within a
disk area of radius 100 m whose center is the same as that of the RRHs.
Each UE has a corresponding potential D2D transmitter that is randomly located
within the distance of 20 m away from the UE.
Each RRH is equipped with two antennas, and each UE is equipped with one antenna.
The channel coefficient of each UE-antenna link consists of the fading related to distance modeled by ${distance^{ - 2}}$, shadow fading of 8 dB, and
small scale fading modeled by ${\cal CN}(0,1)$, while the channel coefficients among UEs are only related to distance.
The maximum transmission power of each RRH is set to 1.5 W, and the constant transmission power of each D2D transmitter is set to
100 mW.
The QoS requirement of each UE is 5 dB.
There are six processors with heterogeneous power consumptions and computing capabilities.
The power consumptions corresponding with these six processors are 21.6 W, 6.4 W,
5 W, 8 W, 12.5 W, and 12.5 W, and their corresponding computing capabilities are
6 MOPTS, 4 MOPTS, 1 MOPTS, 2 MOPTS, 5 MOPTS, and 5 MOPTS.
It is assumed that $P{r_{True,True}}{\rm{ = }}P{r_{False,True}}{\rm{ = }}\rho_m $ for UE $m$,
where $\rho_m$ can be explained as caching service capability of UE $m$'s paired D2D transmitter.
The adopted DQN is a dense NN constructed by an input layer, two hidden layers, and an output layer.
The number of neurons in the input layer is 14, while that in the output layer is 96.
There are 24 neurons in each hidden layer, and ReLu is
utilized as the activation function. All other parameters in the simulation are listed in Table~\ref{table1}.

\begin{table}
\caption{Simulation Parameters}\label{table1}
%\tiny %此处写字体大小控制命令
\small
\centering
%\begin{tabular}{|p{2cm}|p{1 cm}|p{2 cm}|p{1 cm}|p{2 cm}|p{1 cm}|p{2 cm}|p{1 cm}|}
%\begin{tabular}{p{4cm}p{2 cm}p{4 cm}p{2 cm}}
\begin{tabular}{p{2.5 cm}p{1 cm}p{2.5 cm}p{1 cm}}
%\hline
\toprule
\textbf{Parameter} & \textbf{Value}  & \textbf{Parameter} & \textbf{Value} \\
%\hline %画横线
\midrule
 The learning rate of Adam optimizer & 0.0001 & RRH power efficiency & $\frac{1}{40}$ \\
%\hline
\midrule
The capacity of replay memory& 5000 & UE power efficiency & $\frac{1}{20}$ \\
%\hline
\midrule

The number of steps to update target DQN &  480 & Discounted factor & 0.99 \\ %\hline
\midrule

The number of steps to update DQN  & 3 & Noise & $10^{-13}$ W \\
%\hline
\midrule

The number of steps for $\varepsilon$ linearly annealing from 1 to 0.01 & 3000 & Fronthaul transmission power for each UE & 5 W \\ %\hline
\midrule

Batch size for each DQN update  & 32 &The initial steps to populate replay memory by random action selection & 1000 \\
%\hline
\bottomrule
\end{tabular}
\end{table}

\subsection{The Impacts of Learning Parameters}

\begin{figure}[!t]
\center
\includegraphics[width=3.3in]{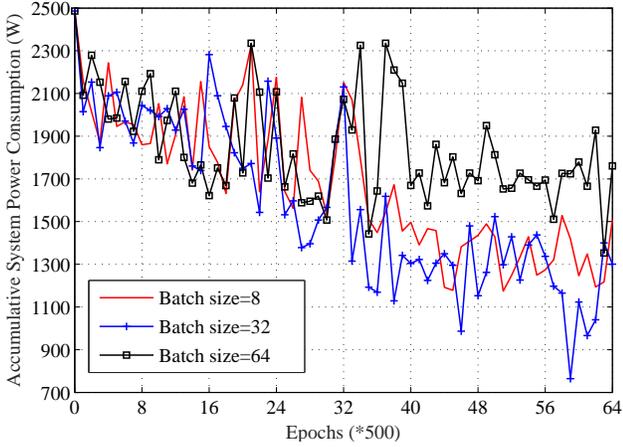}
\caption{The evolution of discounted accumulative system
power consumption under different batch sizes.}\label{batch}
\end{figure}

\begin{figure}[!t]
\center
\includegraphics[width=3.3in]{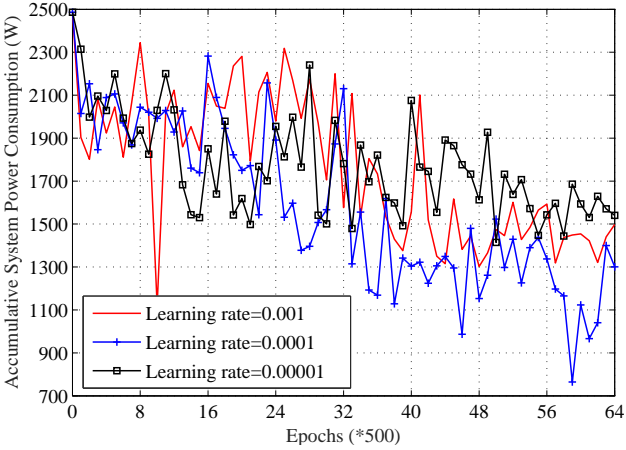}
\caption{The evolution of discounted accumulative system
power consumption under different learning rates.}\label{learning}
\end{figure}

In this subsection, we investigate the impacts of learning rate and batch size on the performance of our proposal
by training DRL with 32000 epochs.
The initial state for each epoch in this section is that all the UEs operate in C-RAN mode with all processors turning on,
and the cache state at each D2D transmitter is \textbf{False}.
From Fig. \ref{batch}, discounted and accumulative system power consumption is evaluated under different batch sizes
with ${\rho _m} = 0.9$, $\forall m$. It can be seen that the performance when batch size is equal to 32 is the best,
whose possible reason can be explained as follows.
With a small batch size, the gradient is only a very rough approximation of the true gradient, and hence
long time can be needed to achieve a good policy.
On the contrary, if the batch size is too large, although the calculated gradient is more accurate,
there is a chance that the learning process is trapped in local optimum.
Under batch size of 32, simulation is conducted to select an appropriate learning rate as shown in Fig. \ref{learning}.
It can be observed that using a too small learning rate 0.00001, the learning process of DRL is slow,
while a larger learning rate 0.001 will result in local optimum. Hence, we select the learning rate as 0.0001.

\subsection{The Impact of Edge Caching Service Capability}

\begin{figure}[!t]
\center
\includegraphics[width=3.3in]{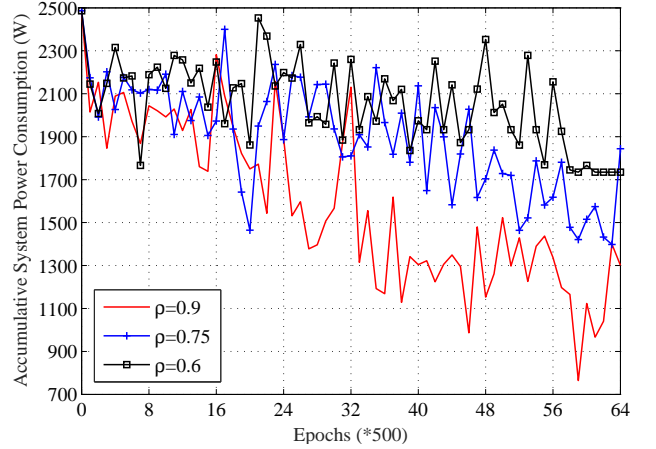}
\caption{The evolution of discounted accumulative system
power consumption under different edge caching service capability.}\label{caching}
\end{figure}

\begin{figure}[!t]
\center
\includegraphics[width=3.3in]{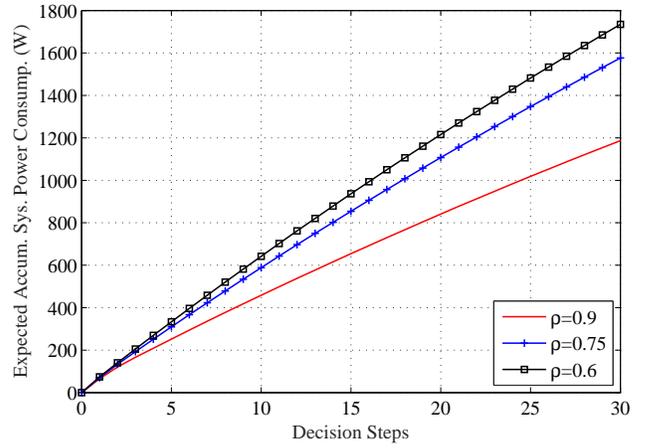}
\caption{The expected discounted accumulative system
power consumption under different edge caching service capability.}\label{cachingave}
\end{figure}

To demonstrate the influence of edge caching on system performance, we let ${\rho _m} = \rho $, $\forall m$,
and vary the value of $\rho$. Fig. \ref{caching} shows the evolution of long term system power consumption
under different $\rho$. It can be seen that a smaller $\rho$ leads to more system power consumption.
This is because when edge caches have poorer service capability,
more UEs need to be served by RRHs, which thus causes larger processor and fronthaul power consumption.
In addition, Fig. \ref{cachingave} is drawn to intuitively show the expectation of long term system performance.
The expected performance for each $\rho$ is estimated by using the corresponding model trained in Fig. \ref{caching} to perform tests over
10000 epochs and then taking the average.

\subsection{The Effectiveness of Integrating Transfer Learning}

\begin{figure}[!t]
\center
\includegraphics[width=3.3in]{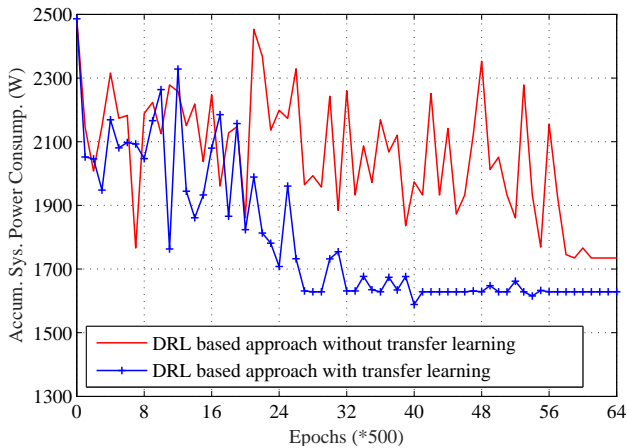}
\caption{The evolution of discounted accumulative system
power consumption with transfer learning when ${\rho _m} = 0.6 $, for all $m$.}\label{transfer0.6}
\end{figure}

\begin{figure}[!t]
\center
\includegraphics[width=3.3in]{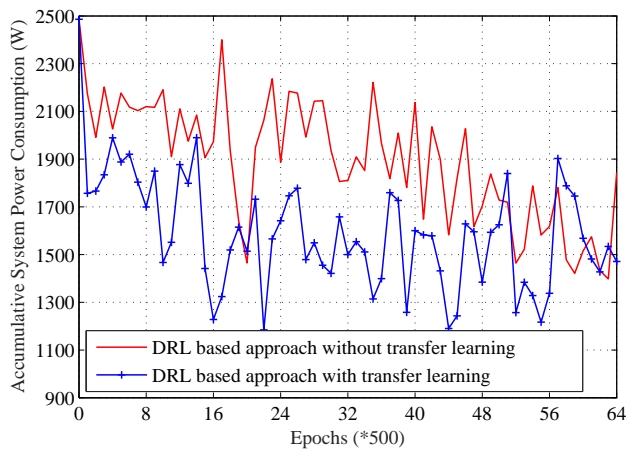}
\caption{The evolution of discounted accumulative system
power consumption with transfer learning when ${\rho _m} = 0.75 $, for all $m$.}\label{transfer0.75}
\end{figure}

To help the DRL model quickly adaptive to new environment where the cache state transition matrix at each D2D transmitter changes,
transfer learning can be adopted, which is expected to accelerate the learning process by transferring
the knowledge learned in a source task to a different but similar task. Since the learned knowledge of the DRL
is stored in the form of connection weights of the DQN, we propose to set the weights of a well-trained DRL model
to another new DRL model to be trained to avoid training from scratch. To verify this idea,
the weights of the DRL model that is trained when ${\rho _m} = 0.9 $, $\forall m$, are used for the weight initialization
of the DRL model to be trained in two different environments with ${\rho _m} = 0.75 $, $\forall m$, and
${\rho _m} = 0.6 $, $\forall m$, respectively.
By the results shown in Fig. \ref{transfer0.6} and Fig. \ref{transfer0.75}, it is observed that
transfer learning can effectively help DRL achieve performance similar to that achieved by training from scratch
but with much less training time.
Nevertheless, transfer learning can lead to negative guidance on the target task when the similarity
between the source task and the target task is low~\cite{transfer}.

\subsection{Performance with Other Baselines}

To verify the superiority of our proposal, the following baseline schemes are adopted in our simulation study:
\begin{itemize}
  \item \textbf{\emph{D2D mode always}}: In this scheme, the controller always progressively makes UEs operate in D2D mode and turns off all the processors.
  \item \textbf{\emph{DRL based, C-RAN mode only}}: In this scheme, all the UEs operate in C-RAN mode, and the controller uses DRL to control the on-off states of processors only.
  \item \textbf{\emph{Q-learning based control}}: In this scheme, the controller controls the UE communication modes and processors' states using the iterative Q-learning based on equation (18).
  \item \textbf{\emph{Random control}}: In this scheme, the controller selects each action with equal probability.
\end{itemize}

\begin{figure}[!t]
\center
\includegraphics[width=3.3in]{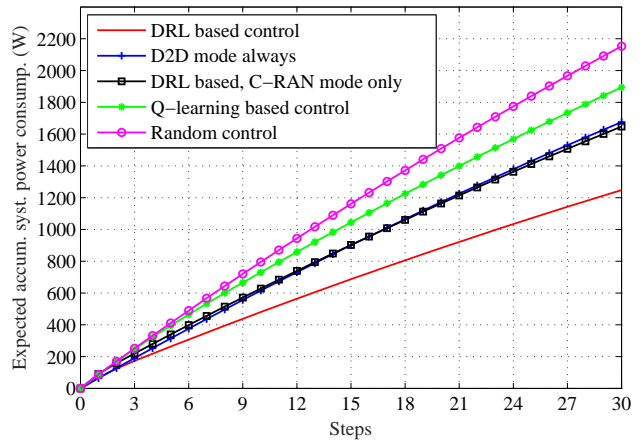}
\caption{Performance comparison with benchmark schemes.}\label{compare}
\end{figure}

Note that in the above comparison baselines, after communication mode selection and processor state control is finished, precoding is optimized using Algorithm 1 if needed,
and the protecting operation still applies to always guarantee the QoS of UEs.
The comparison result is illustrated in Fig. \ref{compare}, where a more general heterogenous caching service capability at each D2D transmitter is considered.
Specifically, we set ${\rho _1} = 0.5$, ${\rho _2} = 0.9$, ${\rho _3} = 0.9$, and ${\rho _4} = 0.5$.
It can be found that our proposal performs the best, which shows its effectiveness on network control facing with dynamic and complex
wireless environment.
Specifically, due to the cache state dynamics and the interference among active D2D links,
the D2D mode always scheme can lead to more frequent D2D communication failure compared with our proposal and hence more frequent
protecting operations. While for the DRL based, C-RAN mode only scheme, although it does not suffer from the dynamic environment since all
the UEs access RRHs, delivering all the traffic via RRH transmission will induce high fronthaul and processor power consumption.
Moreover, compared with Q-learning,
since replay memory helps DRL review the historical interactions
and DQN has the capability of generalizing learned knowledge to new situations, our proposal therefore
achieves better performance with the same number of interactions with the environment.

\section{Conclusion}

In this article, a deep reinforcement learning (DRL) based approach has been developed for a fog radio
access network (F-RAN) to minimize the long-term system power consumption under the dynamics of edge caching states.
Specifically, the network controller can make a quick and intelligent decision on the user equipment (UE) communication modes and
processors' on-off states given the current system state using the well trained DRL model,
and the precoding for UEs in cloud RAN mode is then optimized under per UE quality of service constraints,
per-RRH transmission power constraints, and the computing capability constraint in the cloud based on an
iterative algorithm. By simulations, the impacts of learning rate and batch size have been shown.
Moreover, the impact of edge caching service capability on system power consumption has been demonstrated, and
the superiority of DRL based approach compared with other baselines is significant.
Finally, transfer learning has been integrated with DRL, which can reach performance similar to the case without transfer learning
but needs much less interactions with the environment.
In the future, it is interesting to incorporate power control of device-to-device UEs, subchannel allocation,
as well as fronthaul resource allocation into DRL based resource management to achieve better F-RAN performance
and make the system model more practical.

%
%The proposed approach is fully distributed and the computing burden on each D2D pair is low since
%only simple mathematical operations are executed. Meanwhile, each pair needs to acquire only local channel state information (CSI), and the BBU pool
%only needs to acquire CSI between user equipments and RRHs. The performance gain of enabling D2D in C-RANs has been confirmed by the numerical simulations, and
%the gain depends on the distance between D2D transmitters and RRHs, the fronthaul capacity, as well as the
%centralized signal processing capability of the BBU pool. For future work, it would be interesting to study joint mode selection and resource allocation to optimize the system performance of D2D enabled C-RANs considering both uplink and downlink with advanced distributed optimization techniques.

\end{document}